\documentclass{aa}  
\usepackage{graphicx}
\usepackage{multirow}
\usepackage[dvipsnames]{xcolor}
\usepackage{txfonts }
\usepackage{natbib  }
\usepackage{hyperref}
\begin{document}
  \title{One size fits all: Insights on Extrinsic Thermal Absorption from Similarity of Supernova Remnant Radio Continuum Spectra}


   \author{Mario G. Abadi \inst{1,2}\href{https://orcid.org/0000-0003-3055-6678}{\includegraphics[scale=0.8]{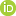}}\thanks{E-mail:
mario.abadi@unc.edu.ar}
\and Gabriela Castelletti\inst{3}\href{https://orcid.org/0009-0002-0134-2064}{\includegraphics[scale=0.8]{orcid.png}}
\and Leonardo Supan\inst{3}\href{https://orcid.org/0000-0002-4309-1403}{\includegraphics[scale=0.8]{orcid.png}}
\and Namir E. Kassim\inst{4}\href{https://orcid.org/0000-0001-8035-4906}{\includegraphics[scale=0.8]{orcid.png}}
\and Joseph W. Lazio\inst{5}
}

\institute{CONICET-Universidad Nacional de C\'{o}rdoba, Instituto de
Astronom\'{i}a Te\'{o}rica y Experimental (IATE),
Laprida 854, X5000BGR,C\'{o}rdoba, Argentina
\and
Observatorio Astron\'{o}mico, Universidad Nacional de C\'{o}rdoba,
Laprida 854, X5000BGR, C\'{o}rdoba, Argentina
             \and 
Instituto de Astronom\'ia y F\'isica del Espacio (IAFE), Ciudad Universitaria - Pabellón 2, Intendente Güiraldes 2160 (C1428EGA), Ciudad Aut\'onoma de Buenos Aires - Argentina
\and
Remote Sensing Division, Naval Research Laboratory, Code 7213, 4555 Overlook Ave SW, Washington, DC, 20375, USA
\and
Jet Propulsion Laboratory, California Institute of Technology, Pasadena, CA 91106, USA
}
   
   \date{Received May XX, 2023; accepted Month XX, 2023}

 
\abstract
{
Typically, integrated radio frequency continuum spectra of supernova remnants (SNRs) exhibit a power-law form due to their synchrotron emission.
In numerous cases, these spectra show an exponential turnover, long assumed due to thermal free-free absorption in the interstellar medium.
We use a compilation of Galactic radio continuum SNR spectra, with and without turnovers, to constrain the distribution of the absorbing ionized gas. We introduce a novel parameterization of SNR spectra in terms of a characteristic frequency $\nu_{\ast}$, which depends both on the absorption turnover frequency and the power-law slope. Normalizing to $\nu_{\ast}$ and to the corresponding flux density, $S_{\ast}$, we demonstrate that the stacked spectra of our sample reveal a similarity in behavior with low scatter (r.m.s. $\sim$15\%), and a unique exponential drop-off fully consistent with the predictions of a free-free absorption process.
Observed SNRs, whether exhibiting spectral turnovers or not, appear to be spatially well mixed in the Galaxy without any evident segregation between them. Moreover, their Galactic distribution does not show a correlation with general properties such as heliocentric distance or Galactic longitude, as might have been expected if the absorption were due to a continuous distribution of ionized gas. However, it naturally arises if the absorbers are discretely distributed, as suggested by early low-frequency observations. Modeling based on \ion{H}{II}  regions tracking Galactic spiral arms successfully reproduces the patchy absorption observed to date. While more extensive statistical datasets should yield more precise spatial models of the absorbing gas distribution, our present conclusion regarding its inhomogeneity will remain robust.}

\titlerunning{Similarity of SNRs}
   \maketitle
%
\section{Introduction}
\label{sec:intro}
Continuum radio emission from shell-type supernova remnants (SNRs) is due to synchrotron emission from relativistic electrons spiraling in compressed circumstellar and interstellar magnetic fields, inferred to be a major source of Galactic  cosmic ray electrons. Their intrinsic nonthermal continuum spectra typically reflect a power-law dependence of the flux density on the frequency, $S\propto \nu^{\alpha}$, the spectral slope of which is shaped by particle acceleration processes associated with their blast wave. Both, linear and nonlinear diffusive shock acceleration models have been proposed to explain the production of the energetic particles and the synchrotron emission (see e.g., \citealt{blandford1987}, \citealt{malkov2001} for reviews).

Free-free thermal absorption from either internal or external ionized gas can impact the integrated radio spectra of SNRs  by causing turnovers at the lowest frequencies ($\nu <$~100 ~MHz). 
In the former case, for shell-type SNRs the absorption is concentrated near the center of the remnant. It is caused by the ionization of cool, unshocked ejecta lying interior to the reverse shock and ionized by its X-ray emission. It is has been detected in the bright, young SNRs Cas~A \citep{kas95,del14,arias+18} and Tycho \citep{arias+19}. Hints of interior absorption attributed to its thermal filaments has also been discerned in the Crab nebula \citep{bie97}. These special cases of interior or {\it intrinsic} thermal absorption in young SNRs remain observationally rare and will no longer be considered in this paper.

The much more common source of turnovers in 
SNR continuum spectra is exterior or {\it extrinsic} absorption 
from ionized gas outside the main boundaries of the remnant and located anywhere along the line of sight. 
Although it is customary to distinguish two different cases, local and non-local, the spectra by themselves do not give any information about the distance between the emitter and the absorber. 
The local case can occur near the periphery of the SNR as the blast wave interacts with the surrounding interstellar medium (ISM) or from foreground \ion{H}{II} regions within the same complex.
On the other hand, the non-local case refers to absorption by unrelated more distant ionized gas. Evidence for spectral turnovers  under 100~MHz with thermal absorption signatures dates to the classic works of \citet{dulk-slee-72,dul75} and \citet{Kassim1989}, but were significantly limited due to poor angular resolution imposed by the Earth's ionosphere on instruments at the time. \citet{Kassim1989} suggested at least some absorption occurs non-locally in  extended ($\sim$100~pc), low-density envelopes (EHEs) associated with normal Galactic \ion{H}{II} regions along the line of sight. EHEs had been inferred earlier from widespread so-called Galactic ridge recombination lines detected at meter wavelengths (325~MHz) \citep{anantha85a}.

Synchrotron emission combined with thermal absorption produce
integrated radio continuum spectra for most Galactic SNRs that can be fit by a power-law plus an exponential turnover at low frequencies ($\nu<100$~MHz). 
Turnover spectra offer crucial insights into the ionized gas properties of the absorbers in our Galaxy. 
In particular, the presence and locations of SNRs with turnover spectra, 
or limits implied by those without turnovers at currently accessible frequencies, offer unique constrains on 
the spatial distribution of this gas, improving our understanding of the interstellar medium. 
However, the use of different reference frequencies ($\nu_{0}$) to parameterize the optical depth ($\tau_{0}$) associated with these turnovers has made it 
awkward 
to compare spectra from different SNRs 
and formulate a clear picture of the underlying absorbing ionized medium
. Therefore, in order to make the comparison of optical depths easier and ensure comparability between published values, it is 
beneficial 
to convert all values to a common reference frequency.
To address this issue, we propose an alternative parameterization of turnover spectra that avoids the need for multiple reference frequencies, streamlining the analysis and facilitating the comparison of different SNR samples.

Early sub-arcminute resolution imaging below 100 MHz of SNRs W49B \citep{lac01} and 3C~391 \citep{bro05-391} began resolving thermal absorption towards SNRs. They foreshadowed emerging arc-second resolution, mJy sensitivity capabilities to expand thermal absorption studies to a larger population of sources for constraining the ionized gas, diagnosing direct interactions, and constraining the relative radial superposition of emitters and absorbers. More recently, \cite{Castelletti2021} 
significantly expanded the sample, by performing a detailed analysis of 9 SNRs showing turnovers on the basis of radio, infrared, and molecular data. 
Four of these have been attributed to  the discrete but non-local extrinsic scenario proposed by \cite{Kassim1989} 
(see "Absorption from extended envelopes of normal \ion{H}{II} regions (EHEs)" in their Table 3). The remaining 5 are far better explained by local extrinsic absorption by foreground ionized gas in the SNR neighbourhood (see "Absorption: Special cases" in their Table 3). Comparing these two tables, the main parameter distinguishing between the non-local and local extrinsic scenarios seems to be the best-fit free-free optical depth $\tau_0$ (at a reference frequency $\nu_0$=74~MHz) in the integrated continuum spectra: a low one, $\tau_0 \sim 0.1$, for the case of external \ion{H}{II} regions versus a high one, $\tau_0 \sim 1 $, in the case of local \ion{H}{II} regions or  
interactions at the ionised interface between SNR blast wave with their host complex. 
The only exception seems to be G39.2$-$0.3 (3C~396), which has a low $\tau_0 \sim 0.063$ but is classified as local rather than non-local absorption. 
Additionally, 
best-fit models including measurements in the low frequency portion of the spectrum are reported in \citet{Kovalenko1994-spec}, and \citet{Kassim1989}.

Here we explore whether sufficient ``trees'' now exist to begin visualizing the ``forest''.  
Is it possible to use the properties of the observed 
(or inferred) 
turnovers, coupled with improving knowledge of SNR distances and \ion{H}{II} region distributions, to constrain the spatial distribution of the absorbers? 
We hope posing this question here can stimulate forthcoming statistical and targeted studies of the seemingly ubiquitous phenomena of thermal absorption towards Galactic SNRs at low frequencies.            
The organization of this paper is  as follows. 
Section~\ref{sec:sample} describes our sample of integrated SNR radio continuum spectra with or without turnovers at low frequencies. 
Section~\ref{sec:analysis} presents our analysis of all spectra in context of external thermal absorption. 
In Sect.~\ref{sec:model} we present our simple Galactic distribution model for absorption of SNR emission. 
The main results and future work are summarized in Sect.~\ref{sec:conclusions}. 

\section{Data collection}
\label{sec:sample}
Our analysis is based on 129 integrated SNR radio continuum spectra. Among them, 57 exhibit a turnover at frequencies below 100~MHz, while the remaining 72 are well fit by simple power law spectra. 
The former group includes 9 SNR spectra taken from \citet{Castelletti2021}, 3 
constructed for this paper, 31 from \citet{Kovalenko1994-spec}, and 14 from \citet{Kassim1989} (see Table~\ref{table:samples}; note that certain sources appear in more than one reference). The group with pure power law spectra comprise 2 sources from \citet{Castelletti2021},
15 remnants whose spectra were 
constructed for this work,  45 from \citet{Kovalenko1994-spec}, and 10 from \citet{Kassim1989} (for reference see the labels in Figs.~\ref{fig:rsn0sto}, \ref{fig:rsn1sto}, and \ref{fig:rsn5sto}).


While constructing the spectra for this study, encompassing both those with and without turnovers, our main goal was to minimize the  typical scatter at the  lower frequencies. 
To achieve this, we conducted new flux density measurements using available images from two resources: the Very Large Array Low-Frequency Sky Survey Redux (VLSSr, at 74 MHz, resolution 75$^{\prime\prime}$; \citealt{Lane+04}) and the Galactic and Extragalactic All-sky Murchison Widefield Array survey (GLEAM, at 88, 118, 155, and 200~MHz, resolutions from $\sim$$6^{\prime}$ to $\sim$$2^{\prime}$ \citealt{hurley19}). 
These fluxes are critical for constraining the low-frequency portion of the spectra, which 
for many years have remained problematic due to the relatively poor angular resolution and sensitivity of legacy instruments, such as Culgoora (80~MHz, \citealt{slee-higgins-73,slee-higgins-75,Slee-77}), Clark Lake TPT (30.9~MHz, \citealt{kassim-88}), and the Pushchino telescopes (83~MHz, \citealt{Kovalenko1994}).\footnote{Additional flux density measurements below 100~MHz, which were included in our spectral analysis and obtained from the literature, are as follows: 10~MHz \citet{Bridle1968}, 19~MHz \citet{Rishbeth1958}, 10-25~MHz  \citet{Braude1969,Braude1979},  22~MHz \citet{Roger1969,Roger1986}, 26~MHz \citet{Erickson1965,Viner1975}, 29.9~MHz \citet{Jones1974}, 38~MHz \citet{Williams66}, and 86~MHz \citet{Mills1957,Mills1958,Mills1960}. 
}
It is worth noting that at the time of our analysis, no images below 100 MHz were available for the SNRs in our sample from surveys conducted with modern instruments like the LOw Frequency ARray (LOFAR) or the Long Wavelength Array (LWA). 
At higher frequencies, depending on the position of the SNR in the sky, the spectra incorporate new flux density measurements from the Southern Galactic Plane Survey (SGPS, \citealt{Mcclure+05}) at 1420~MHz, the S-band Polarisation All Sky Survey (S-PASS, \citealt{Carretti+19}) at 2303~MHz, and the Parkes 6~cm survey at 5000~MHz \citep{Haynes+78}. 
All our new flux estimates have been combined with previous measurements from the literature. The criteria adopted for collecting the data for the new spectra are consistent with those  reported in \citet{Castelletti2021}. This includes encompassing only measurements with error estimates less than 30\%, ensuring there are no significant deviations from the best-fit model, using interferometer data that adequately recover short spacings information at frequencies above 1000~MHz, and single-dish data with appropriate resolution to prevent flux density overestimates due to high confusion levels. In total, considering the spectra from \citet{Castelletti2021} and those constructed for this work,  we have collected around 570 flux density measurements. 
The improvements in the accuracy of the radio spectral index determinations for the SNRs in our sample 
are about one order of magnitude when compared to previously published estimates. 


\section{Analysis}

%
%
%
\begin{figure}
       \includegraphics[width=0.45\textwidth]{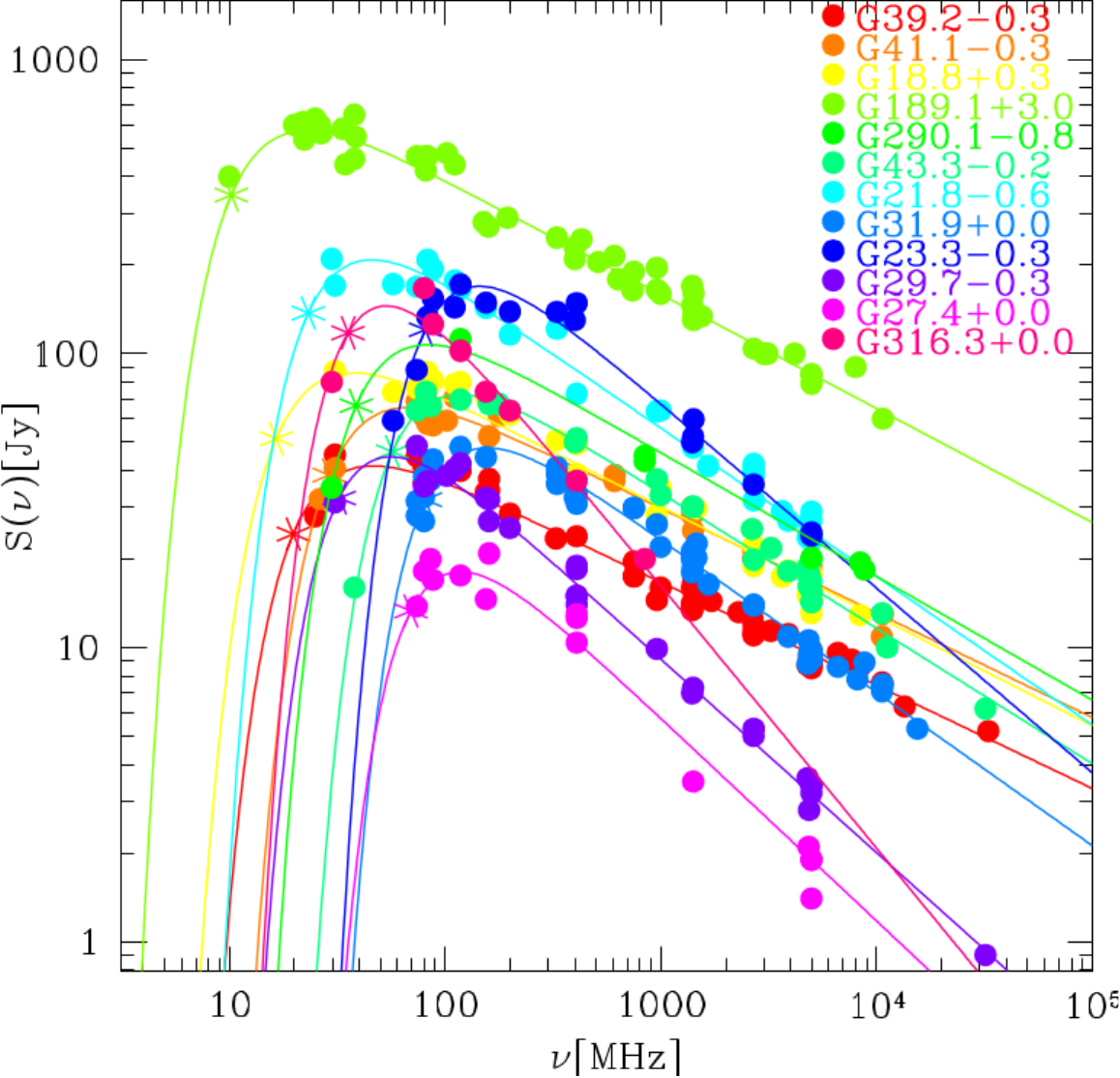}
    \caption{Radio flux density measurements (colored filled circles) as a function of frequency for  
    12  SNRs spectra with turnovers (see labels). This sample is build up by 9 SNR spectra reported in \citet{Castelletti2021} plus 3 new spectra for the SNRs G189.1+3.0, G290.1$-$0.8, and G316.3+0.0, which were specifically constructed for this work. The solid colored lines shows each corresponding best fit given by Eq.~\ref{equation:eq2} with parameters included in our Table~\ref{table:samples}. The asterisk symbol indicates the individual characteristic frequency $\nu_{\ast}$.}
    \label{fig:rsn0}
\end{figure}

\label{sec:analysis}
SNRs that exhibit thermal turnovers are typically characterized by fitting the integrated continuum flux density $S(\nu)$  at a frequency $\nu$ using a power-law plus exponential cutoff equation:
\begin{equation}
S(\nu)=S_0 \Big( \frac{\nu}{\nu_0} \Big) ^{\alpha} \mathrm{exp} \Big[ -\tau_0 \Big(\frac{\nu}{\nu_0}\Big)^{-2.1} \Big],
\label{equation:eq1}
\end{equation}
where $\alpha$ is the power-law spectral index, $S_0$ is a flux density normalization,   and $\tau_0$ is the free-free optical depth, both measured at a reference frequency $\nu_0$. The power-law term in this equation reflects the intrinsic synchrotron emission of the SNR, while the exponential term accounts for absorption due to ionized gas along the line of sight. However, because authors adopt different reference frequencies, typically related to their specific observing frequencies, it is often necessary to apply the scaling : $\nu'_0/\nu_0=(\tau_0/\tau'_0)^{1/2.1}$ 
and $S'_0/S_0=(\nu'_0/\nu_0)^{\alpha}$ where primed variables correspond to a reference frequency $\nu'_0$. 
Some authors (e.g., \citealt{Castelletti2021}) use the same reference frequency ($\nu_0=74$~MHz) for both physical process, i.e. in the power-law and in the exponential; others (e.g., \citealt{Kassim1989}) present their data using two different reference frequencies, one for the emission $\nu_0=408$~MHz and one for the absorption $\nu_0=30.9$~MHz. 
\begin{figure}
\includegraphics[width=0.45\textwidth]{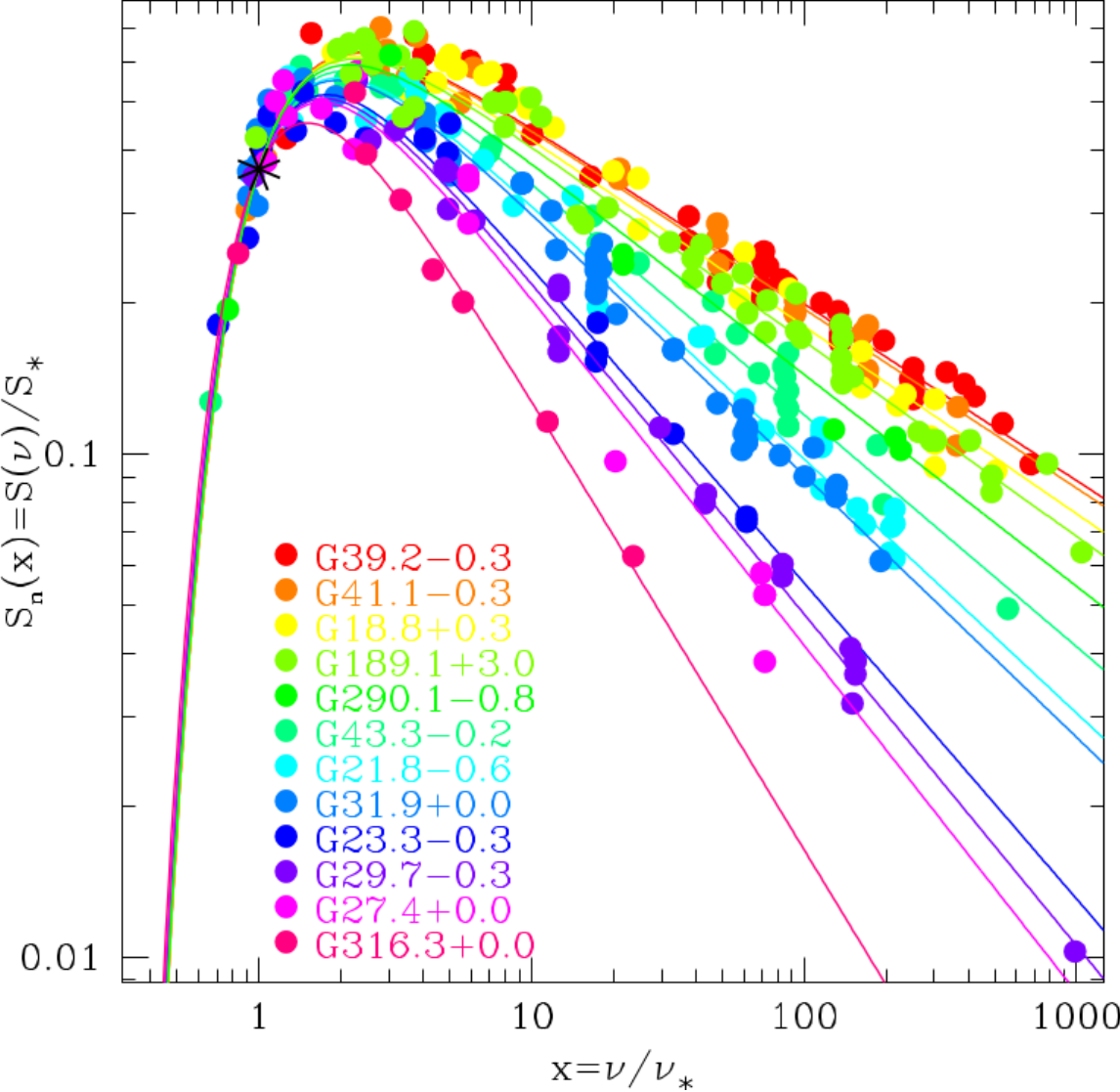}
\caption{Normalized radio flux density $S_{n}(x)=S(\nu)/S_{\ast}=x^{\alpha}\mathrm{exp}(-x^{-2.1})$ measurements (colored filled circles) as a function of the normalized frequency $x={\nu}/{\nu_{\ast}}$ for 12 SNRs spectra with turnover (see labels). 
This sample comprises 9 SNR spectra reported in \citet{Castelletti2021}, with the addition of  3 new spectra for the SNRs G189.1+3.0, G290.1$-$0.8, and G316.3+0.0 included for this work.
Solid lines show  their corresponding fit 
with parameters presented in our Table~\ref{table:samples}. This normalized form sorts SNRs spectra in increasing order accordingly to their slope from bottom $\alpha=-0.89$ to top $\alpha=-0.35$. The black asterisk symbol marks the characteristic frequency $\nu_{\ast}$ and highlight how the normalization is done.
}
\label{fig:rsn1}
\end{figure}
Moreover, others (e.g., \citealt{Kovalenko1994-spec}) use a unique reference frequency for absorption $\nu_0=100$~MHz but different ones for each individual SNR emission over a broad range from $\nu_0= 175$~MHz to $\nu_0=6000$~MHz (see Table~\ref{table:samples}). Although it is straightforward to convert values from one frequency to the other, the discordant parametrization can confuse comparison. To avoid such complications we re-write Eq.~\ref{equation:eq1} independent of the reference frequency as: 
\begin{equation}
S(\nu)=S_{\ast} \Big( \frac{\nu}{\nu_{\ast}} \Big) ^{\alpha} \mathrm{exp} \Big[ -\Big(\frac{\nu}{\nu_{\ast}}\Big)^{-2.1} \Big],
\label{equation:eq2}
\end{equation}
where $\nu_{\ast}$ is a characteristic frequency and $S_{\ast}$ is a characteristic flux density. Equation~\ref{equation:eq2} is identical to Eq.~\ref{equation:eq1} but expressed by removing the explicit dependency of $\nu_0$, $\tau_0$, and $S_0$, i.e., independent of the adopted reference frequency $\nu_0$. Notice that $\nu_{\ast}=\tau_0^{1/2.1} \nu_0$ and $S_{\ast}=S_0(\nu_{\ast}/{\nu_0})^{\alpha}$. Computing these characteristic parameters using any other reference frequency gives the same values of~$\nu_{\ast}$ and~$S_{\ast}$. 
Moreover, the physical meaning of these two parameters is straightforward: $\nu_{\ast}$ is related to the position of the turnover frequency $\nu_{\rm to}$, i.e., where $S(\nu)$ has its maximum, and $S_{\ast}$ is the spectrum normalization height. The turnover frequency can be computed by deriving the flux density and equalizing it to zero to yield 
$\nu_{\rm to}/\nu_{\ast}= (-2.1/\alpha)^{1/2.1}$. 
Then, the characteristic frequency $\nu_{\ast}$ is determined by both the turnover frequency $\nu_{\rm to}$ and the power-law slope $\alpha$. Only in the critical case where $\alpha=-2.1$ we have $\nu_{\ast}=\nu_{\rm to}$ while for $\alpha > (<) -2.1$, we have $\nu_{\ast} < (>) \nu_{\rm to}$. For a fiducial value of $\alpha \sim -0.5$  we obtain $\nu_{\ast} \sim \nu_{\rm to}/2$, while for typical values of SNRs power-law slopes $-0.89 \lesssim \alpha \lesssim -0.35$ (see Table~\ref{table:samples}) we have $1.5 \lesssim \nu_{\rm to}/\nu_{\ast} \lesssim 2.5$. It is  worth noting that while Eq.~\ref{equation:eq1} has been widely adopted by many authors, to the best of our knowledge, the simple parametrization given by Eq.~\ref{equation:eq2} has not been proposed before.
An additional advantage of this parametrization in terms of $\nu_{\ast}$ is that low values of $\nu_{\ast}$ indicate low $\nu_{\rm to}$ or lower frequency turnovers, while high $\nu_{\ast}$ indicate high $\nu_{\rm to}$ with absorption commencing at higher frequencies.  

To illustrate our approach to constrain the properties of the absorbing gas, in the next two subsections we apply these ideas to a sample of 129 SNRs spectra with and without observed turnovers. For the latter group, we utilize upper limits for $\nu_{\ast}$ based on the lowest frequency measured.
\begin{figure}
       \includegraphics[width=0.45\textwidth]{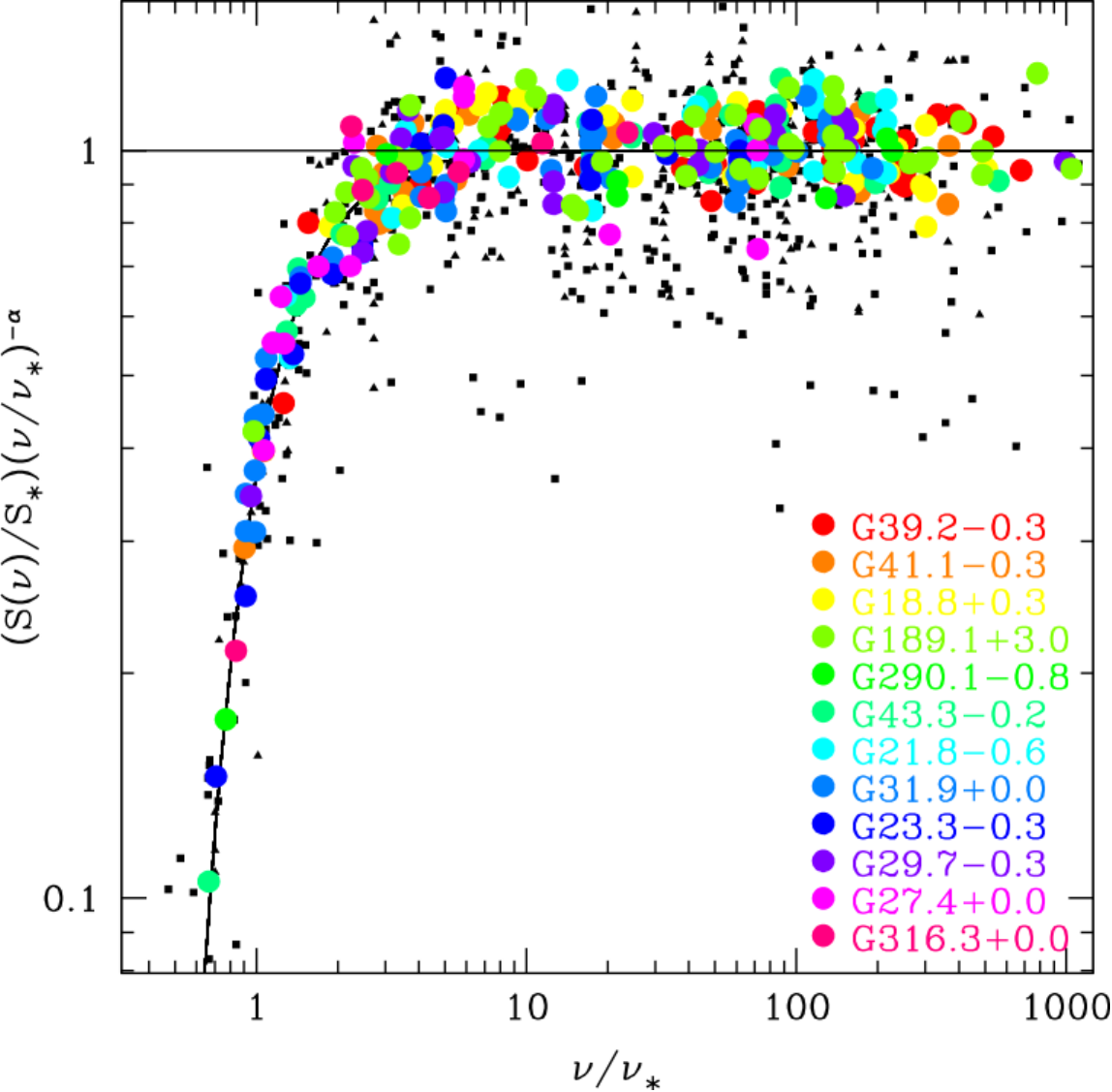}
    \caption{Normalized radio flux density measurement as a function of frequency 
    for 9 SNR spectra presented in \citet{Castelletti2021} plus 3 new spectra for the SNRs G189.1+3.0, G290.1$-$0.8, and G316.3+0.0, constructed for this work (colored filled circles), along with data from \citet{Kovalenko1994-spec} (black filled squares) and \citet{Kassim1989} (black filled triangles). 
    In order to highlight the drop-off behavior, the data are normalized to the power-law emission $E(x)=x^{\alpha}$, where $x=\nu/\nu_{\ast}$.
    The curved solid line is the normalized absorption $A(x)=\mathrm{exp}(-x^{-2.1})$. The total r.m.s. of the colored circles is less than $11\%$. 
    At $\nu \sim 2.75 \nu_{\ast}$ approximately 50\% of the emission is absorbed.}
    \label{fig:rsn5}
\end{figure}


\subsection{ 
SNRs with spectral turnovers}
In Fig.~\ref{fig:rsn0} we display the measurements of 12 SNRs with turnovers, 9 of them reported by \cite{Castelletti2021} plus 3 
constructed for this work.
Accordingly, Fig.~\ref{fig:rsn0} illustrates the complex superposition of spectra reflecting the relatively large range in spectral index ($-0.89 \lesssim \alpha \lesssim -0.35$) 
and optical depth ($0.016 \lesssim \tau_0 \lesssim 1.21$ at $\nu_0=74$~MHz) listed in Table~\ref{table:samples}. 

In Fig.~\ref{fig:rsn1} we show the first advantage of the new parametrization offered through Eq.~\ref{equation:eq2}. Here, we plot the normalized flux density $S_n(x)=S(\nu)/S_{\ast}$ as a function of the normalized frequency $x=\nu/\nu_{\ast}$ for the same 
12
SNRs plotted in Fig.~\ref{fig:rsn0}. 
The dynamic range of 
$\sim 700$ 
in $S(\nu)$ seen in Fig.~\ref{fig:rsn0} is reduced by at least a factor 
$\sim 10$ 
using the normalized form also showing a significantly smaller scatter. Moreover, the observational data at low  frequencies, i.e. lower than the characteristic frequency $\nu<\nu_{\ast}$ of the 
12 
SNRs when stacked together, are well fit by a unique drop off curve $A(x)=\mathrm{exp}(-x^{-2.1})$ that is more clearly traced than in the individual curves.  
At high frequencies, i.e. higher than the characteristic frequency $\nu>\nu_{\ast}$, the solid curves differ only due to their slopes $\alpha$. Power-law emissions are ordered by their slope from a steep 
$\alpha = -0.89$ 
value for 
G316.3+0.0
( 
bright pink
filled circles and solid curve) 
to a flat $\alpha =-0.35$ for G39.2$-$0.3 (red filled circles and solid curve). 

In Fig.~\ref{fig:rsn5} we applied 
an additional normalization step by dividing the flux density $S_{n}(x)$ by the power-law emission 
factor 
of each SNR, $E(x)=x^{\alpha}$.
Another advantage of the new normalization is that any SNR spectrum can be fit by a unique absorption equation $A(x)=\mathrm{exp}(-x^{-2.1})$, as depicted by the solid black curve.
   A  reduced r.m.s. scatter of $\sim 11\%$ is apparent for these 
   12 SNRs 
   stacked together indicating the quality of the fits. 
   To enlarge our sample, we have also compiled spectra with a low-frequency turnover of 14 SNRs reported in \citet{Kassim1989} and 31 from \citet{Kovalenko1994-spec}. These dataset are represented by filled small black squares and triangles, respectively. 
 The inclusion of these two additional samples reinforces our results showing a similar trend, albeit with a higher r.m.s. scatter.

\subsection{Pure power-law SNRs spectra}
\label{sec:section2.2}
Pure power-law spectra can be interpreted as absorbed spectra as well, but absorbed at such low frequencies that the turnover has not been observationally detected yet. For example,  if the turnover frequency is at $\nu_{\ast}=30$~MHz, but there are no quality measurements at comparable frequencies, the  spectrum is usually fit by a pure power-law.  Indeed, if a turnover has not been detected, the lowest measured frequency  $\nu_\mathrm{low}$ of a spectrum can be considered as an upper limit for the undetected turnover frequency. We therefore expanded our sample of turnover spectra by assigning a turnover to the pure power law fits, allowing a maximum of a 10\% difference between the extrapolated power-law and the absorbed fit at the lowest frequency. 
Using Eq.~\ref{equation:eq2}, we have $\nu_{\ast}=\nu_\mathrm{low}(\log f)^{1/2.1}$ where $f$ is a parameter that fixes the ratio between the flux density power-law and the turnover fit at $\nu_\mathrm{low}$. We have adopted a value of $f=1.1$, which means that the difference is 10\% yielding $\nu_{\ast}\sim0.33\nu_{\mathrm{low}}$. 
 
Figs.~\ref{fig:rsn0sto}, ~\ref{fig:rsn1sto}, and ~\ref{fig:rsn5sto} are analogous to Figs.~\ref{fig:rsn0}, ~\ref{fig:rsn1}, and ~\ref{fig:rsn5}, respectively,  but constructed for SNRs whose integrated radio continuum spectra are fitted by pure power laws. Two of the plotted spectra are taken from \citet{Castelletti2021}, and the other 15 were constructed for this work as explained in Sect.~\ref{sec:sample}. In Fig.~\ref{fig:rsn0sto} colored solid lines  are the corresponding pure power-law best fit to the data points of each SNR, while dotted lines are the putative turnover spectra given by Eq.~\ref{equation:eq2}. The asterisk symbols indicate the maximum characteristic frequency $\nu_{\ast}$ allowed by the lowest frequency observed $\nu_\mathrm{low}$. Again, the large dynamic range in measured flux densities of $\sim 300$ is reduced by a factor $\sim 7$ when the normalized form is adopted (see Fig.~\ref{fig:rsn1sto}). A final normalization to the individual power law slope $\alpha$ shows good quality fits with r.m.s. $\sim 15\%$ as seen in the colored open circles of Fig.~\ref{fig:rsn5sto}. These fits are better quality than the  pure power-law samples of \cite{Kovalenko1994-spec} and \cite{Kassim1989} depicted by open small black squares and triangles, respectively.

\subsection{Characteristic frequency $\nu_{\ast}$ distribution} 
In Fig.~\ref{fig:histo} we show as a filled histogram the distribution of $\nu_{\ast}$ values,  computed for our three samples (\citet{Castelletti2021} plus this work, \citet{Kovalenko1994-spec}, and \citet{Kassim1989} listed in Table~\ref{table:samples}) ranging from $\sim$10~MHz to 100~MHz. 
\begin{figure}
       \includegraphics[width=0.45\textwidth]{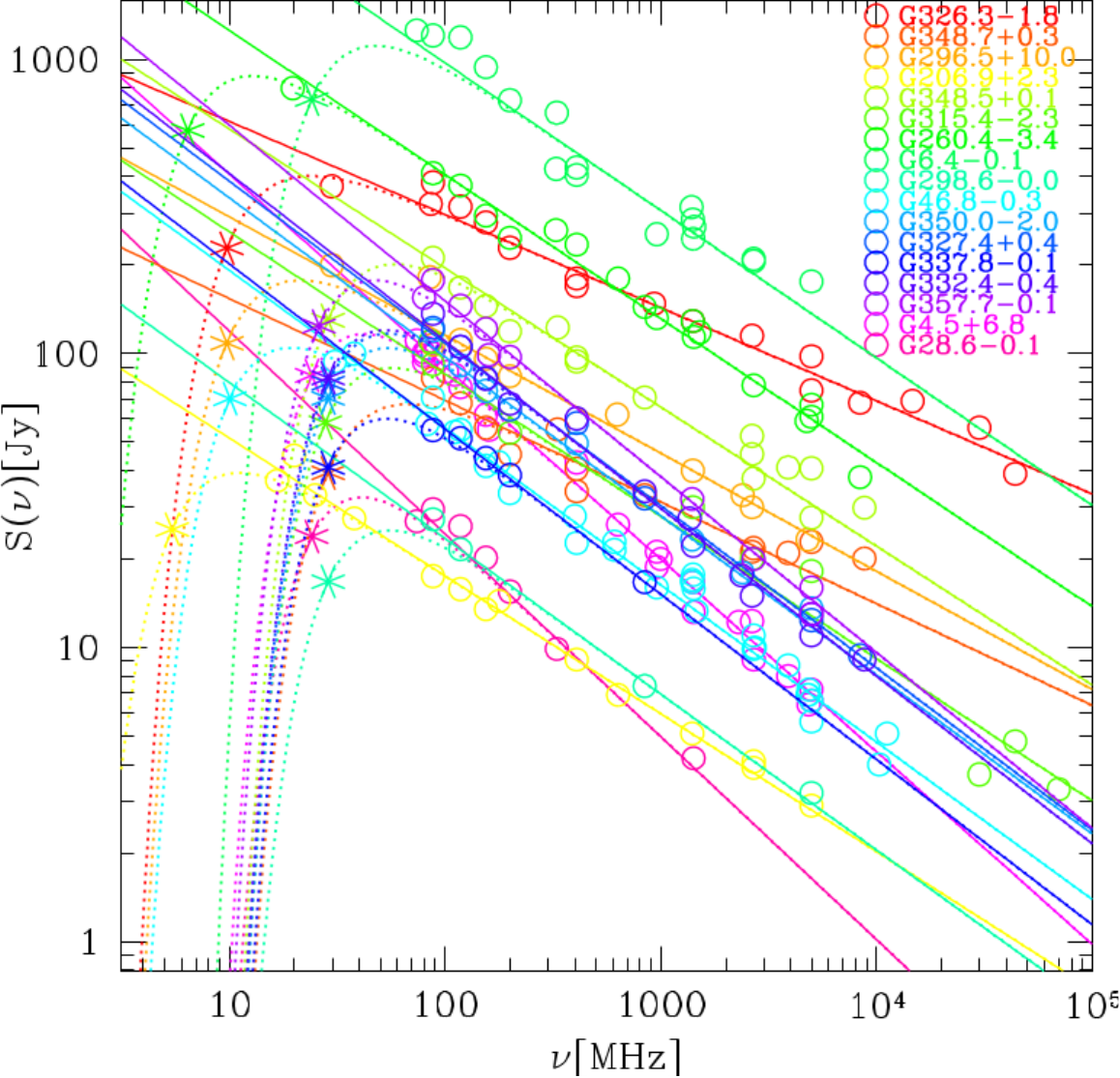}
    \caption{Radio flux density measurements (colored open circles) as a function of frequency for 17 SNRs spectra without turnovers (see labels). This sample includes the spectra of G4.5+6.8 and G28.6$-$0.1 SNRs published in \citet{Castelletti2021} plus 15 new ones constructed for this work. The solid colored lines shows each corresponding power-law best fit while the dotted ones corresponds to  
    Eq.~\ref{equation:eq2} with a characteristic frequency $\nu_{\ast}$ (see asterisk symbol) assigned as an upper limit according to the lowest measured frequency $\nu_\mathrm{low}$.}
    \label{fig:rsn0sto}
\end{figure}
In the upper panel, $\nu_{\ast} \sim 40$~MHz separates two populations   with low ($\tau \sim 0.1$) 
and high 
($\tau  \sim 1$)
optical depths reported 
in Table~3 of \cite{Castelletti2021}; these two subsamples were  labeled as ``Absorption from extended envelopes'' (non-local) and 
``Absorption: Special cases'' (local), respectively. The classification was made by additional information obtained from low-frequency radio recombination lines, infrared fine-structure, and molecular line emission to distinguish between spectral turnovers attributed to ionized material in the ISM unrelated to the SNR (non-local) and 
an ionized component in the immediate SNR's surroundings (local). 
These low ($\nu_{\ast} \lesssim 40$~MHz) and high ($\nu_{\ast} \gtrsim 40$~MHz) populations might hint at non-local vs. local absorbers in the other two samples, \cite{Kassim1989} and \cite{Kovalenko1994-spec}, shown in the middle and bottom panels of Fig.~\ref{fig:histo}.
However, due to the incompleteness of our samples, it is not possible to definitively determine yet whether these populations represent truly distinct groups or merely the low and high tails of a single population with intrinsic dispersion. 
The open histograms show the characteristic frequency $\nu_{\ast}$ of  power-law spectra assigned according to their observed lowest frequency $\nu_\mathrm{low}$. 
In the upper panel, the observed peak at $\nu_{\ast}\sim30$ MHz is a direct consequence of the GLEAM survey cutoff frequency at $\nu_\mathrm{low} \sim 88$~MHz (see asterisk symbols in Fig.~\ref{fig:rsn0sto}). At $\nu_{\ast}\sim10$~MHz we have 
measurements obtained at lower frequencies by the Clark Lake TPT (30.9~MHz), Culgoora (80~MHz), and Pushchino (83~MHz) telescopes (see Sect.~\ref{sec:sample}). 

\begin{figure}
       \includegraphics[width=0.45\textwidth]{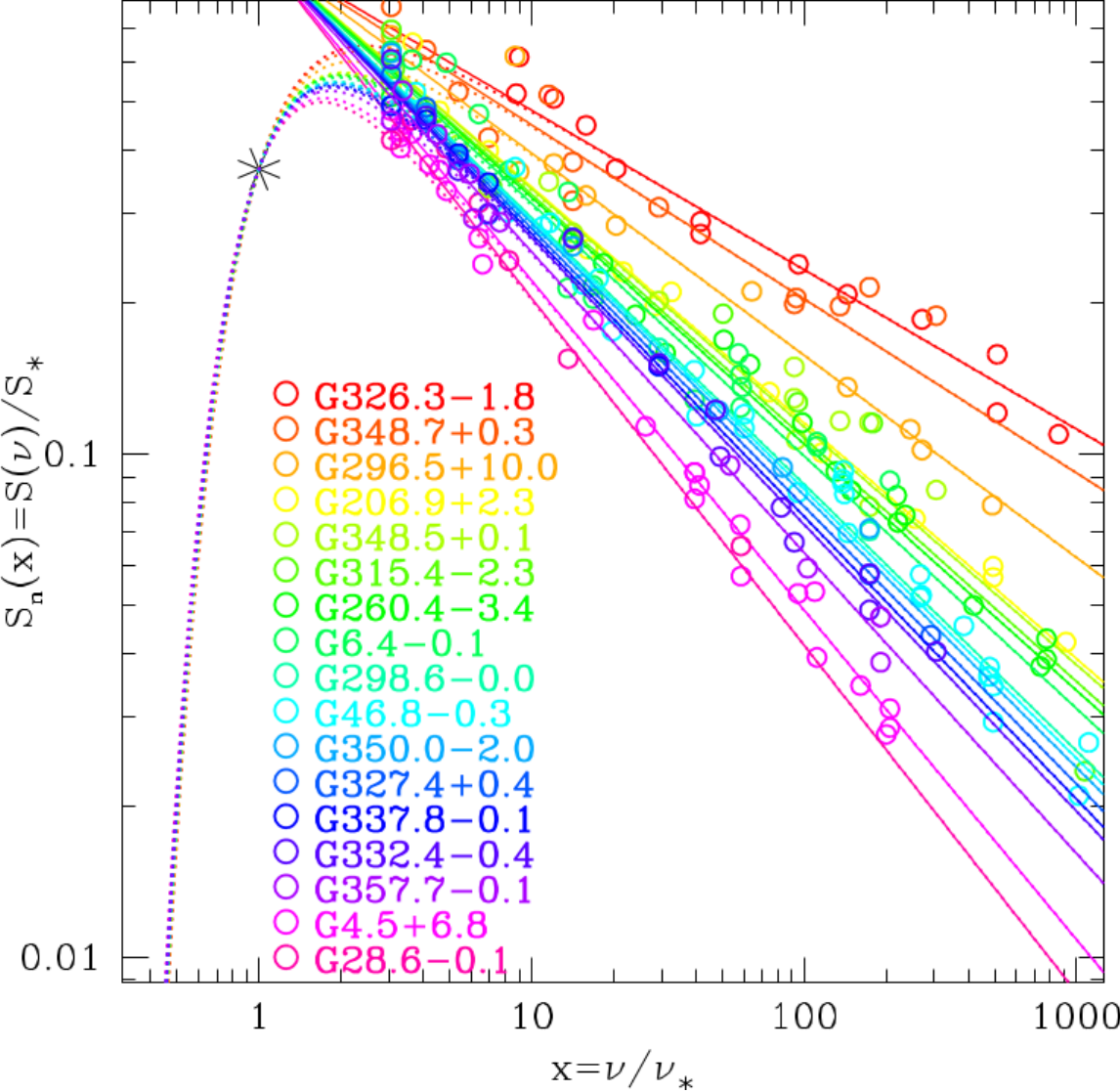}
    \caption{Normalized radio flux density $S_{n}(x)=S(\nu)/S_{\ast}=x^{\alpha}\mathrm{exp}(-x^{-2.1})$ measurements (colored open circles) as a function of the normalized frequency $x={\nu}/{\nu_{\ast}}$ for 17 SNRs spectra without low frequency turnover (see labels). 
    This sample includes the spectra of G4.5+6.8 and G28.6$-$0.1 SNRs published in \citet{Castelletti2021} plus 15 new collected for this work. Extended lines show their power-law fit while dotted ones  correspond to Eq.~\ref{equation:eq2} with characteristic frequency $\nu_{\ast}$ (asterisk symbol) assigned as an upper limit according to the lowest measured frequency $\nu_{\mathrm{low}}$. This normalized form sorts SNRs spectra in increasing order accordingly to their slope from bottom $\alpha=-0.69$ to top $\alpha=-0.32$. The black asterisk symbol marks the characteristic frequency $\nu_{\ast}$ and highlights how the normalization is done.} 
    \label{fig:rsn1sto}
\end{figure}

\begin{figure}
    \includegraphics[width=0.45\textwidth]{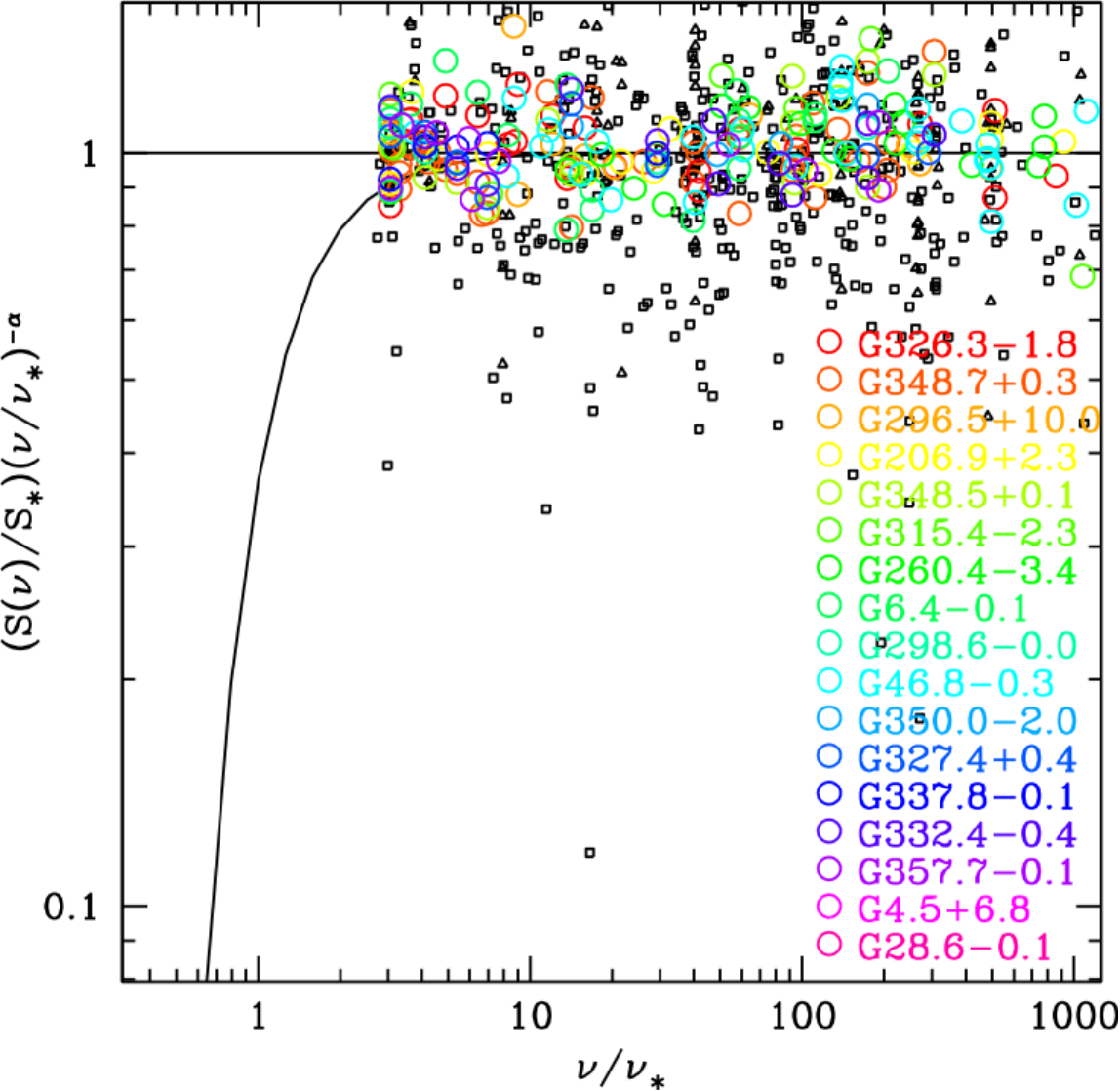}
    \caption{Normalized radio flux density measurements as a function of frequency  for the SNRs G4.5+6.8 and G28.6$-$0.1 presented in  \citet{Castelletti2021}  plus 15 new SNRs spectra constructed for this work (colored open circles). These datasets are complemented by flux density  estimates  from  \citet{Kovalenko1994-spec} (black open squares) and \citet{Kassim1989} (black open triangles). 
    In order to highlight the drop-off behavior, the data are normalized to the power-law emission $E(x)=x^{\alpha}$, where $x=\nu/\nu_{\ast}$. 
    The curved solid line is the normalized absorption $A(x)=\mathrm{exp}(-x^{-2.1})$. The total r.m.s. of the colored open circles is less than $15\%$.
    }
    \label{fig:rsn5sto}
\end{figure}

\begin{figure}
       \includegraphics[width=0.45\textwidth]{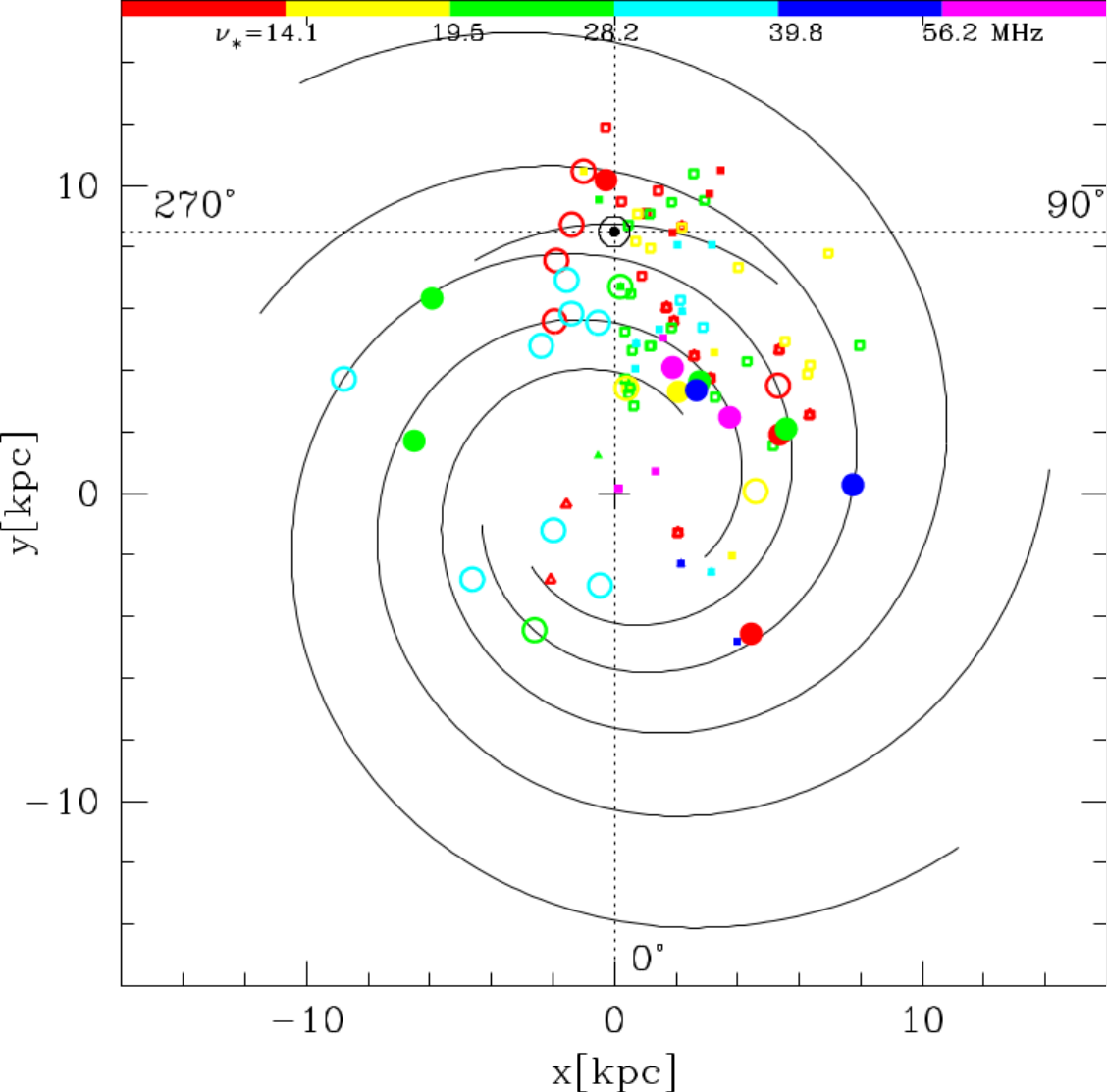}
    \caption{Projected spatial distribution onto the Galactic  plane of SNRs in the three samples analyzed here 
    with distances taken from \citet{ranasinghe2022} and colored according to  ${\log}(\nu_{\ast}/\rm{MHz})$ using the color bar code. Filled symbols show the location of SNRs with turnover spectra, while the open symbols indicate those sources not showing turnovers. Circles correspond to \citet{Castelletti2021} plus this work data, squares correspond to \citet{Kovalenko1994-spec}, and triangles to \citet{Kassim1989}. 
    As a guidance the solid black lines show the position of the spiral arms taken from the  \citet{Hou&Han2014} model. The central plus symbol indicates the position of the Galactic  center, the solar symbol shows the Sun's position, and the dotted lines the Galactic quadrants. The colored points do not show any strong dependence on  heliocentric distance or Galactic longitude (see Fig.~\ref{fig:nusdisall} and\ref{fig:nuslonall}).}
    \label{fig:spiralarms}
\end{figure}

Notice that the upper color bar indicates the color coding adopted in Fig.~\ref{fig:spiralarms} and Fig.~\ref{fig:sa} based on  $\log(\nu_{\ast}/\mathrm{MHz})$ value instead of using the SNRs' names, as done in 
Fig.~\ref{fig:rsn0} through Fig.~\ref{fig:rsn5sto}, as well as in Fig.~\ref{fig:nusdisall} and Fig.~\ref{fig:nuslonall}. 
On the other hand, we have also searched for, and did not find, any significant correlations between the three fitted parameters $S_{\ast}$, $\nu_{\ast}$, and $\alpha$. This result indicates that the emission and absorption processes in our limited SNR sample are independent of each other.
A useful and direct application of the characteristic frequency $\nu_{\ast}$ parametrization is the possibility to provide insights concerning the properties of the absorbing thermal gas. A complete mapping of the $\nu_{\ast}$ distribution in the Galaxy should offer strong and unique constraints on the distribution of the ionized absorbing gas. 
Despite the incomplete and non-uniform nature of current surveys on SNRs, we have utilized our compiled sample to extract fundamental and widespread characteristics from the presently available data. Additionally, we have compared our findings with ad-hoc constructed models in order to draw initial conclusions regarding the absorber population. 
In Fig.~\ref{fig:spiralarms} we show the projected spatial distribution onto the Galactic plane for all the SNRs with turnovers (filled symbols) in our three samples.
We have also included SNRs  from 
\cite{Castelletti2021} plus this work, \cite{Kovalenko1994-spec}, and \cite{Kassim1989} 
whose spectra are well-fitted by a pure power-law (open symbols) colored according to their upper limit characteristic frequency assigned (see Sect.~\ref{sec:section2.2} and Fig.~\ref{fig:rsn0sto}). 
At first glance, it does not appear that there is a significant segregation between SNRs with and without spectral turnovers, or an apparent gradient in $\nu_{\ast}$ with global properties like heliocentric distance or Galactic  longitude. 
This qualitative result indicates that the absorbing media is probably made of discrete ionized regions rather than a continuous gas distribution. Likewise,
the absence of spectral turnover in some distant SNRs, coupled with significant absorption in some nearby ones, was first interpreted by \citet{Kassim1989} as a patchy distribution of low-frequency absorbing gas, and not caused by a broadly distributed component of the ISM. Here, we test this original idea by adding the spectra presented by \citet{Castelletti2021}, those constructed for this work, along with those included in \citet{Kovalenko1994-spec}.

To facilitate the interpretation of these observational findings, in the following section we developed a two-dimensional toy model that simulates the power-law emission from SNRs and takes into account the absorption caused by intervening ionizing gas.

\section{Absorption Model} \label{sec:model}
We assumed that the SNRs are 
distributed on the Galactic  plane following an exponential surface number density profile: $\Sigma(R)=\Sigma_0 \exp(-R/R_d)$, where $\Sigma_0$ is the central surface density and $R_d$ is the scale-length.  We tied $\Sigma_0$ to a mock assumption of 20,000 SNRs sufficiently large to mimic the statistical behavior of a generic population of remnants distributed on the Galactic  plane with $R_d=3.5$~kpc following the Milky Way stellar scale-length. We started by attributing the absorption to ionized gas characterized by an emission measure (see Eq.~\ref{EM} and discussion below) predicted by the continuum model NE2001, at fixed temperature
\citep{cordes2002}. We found that the NE2001 model predicts a correlation between absorption and distance, a correlation that is not observed. We attribute the discrepancy to the fact that the NE2001 model assumes a nearly continuous distribution of ionized gas, together with its reliance on measurements of relatively nearby pulsars (PSRs). 
Recognizing that \ion{H}{II} regions are discrete structures, we then considered a model in which a set of discrete absorbing regions, representing \ion{H}{II} regions or \ion{H}{II} region complexes, are distributed throughout the disk of the Galaxy.

We place clumps of ionized gas preferentially along the Galactic  spiral arms following the \cite{Hou&Han2014} four-arm logarithmic model (see their Table 1). 
 Clumps 
are circular with radius $r$ drawn from a Maxwell-Boltzmann probability density distribution:

\begin{figure}
       \includegraphics[width=0.45\textwidth]{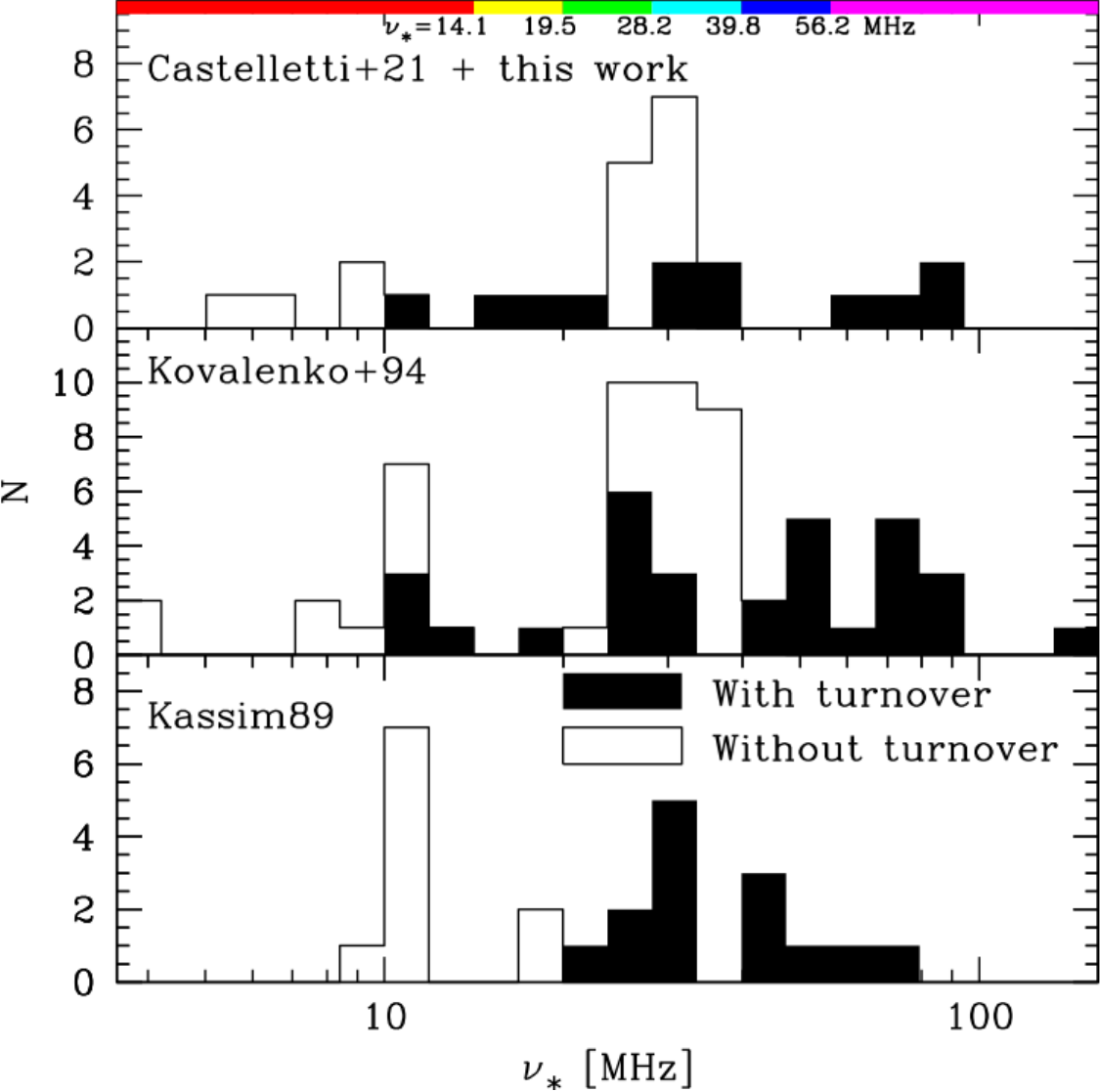}
    \caption{Histograms showing the distribution of characteristic frequencies $\nu_{\ast}$ for the three samples analyzed here. Each panel corresponds to a different sample, as indicated by the labels.
    The filled histograms show the distribution of $\nu_{\ast}$ for SNRs spectra with turnovers, while the open ones depict the estimated upper limit of $\nu_{\ast}$  based on the lowest measured frequency $\nu_{\mathrm{low}}$ for  spectra without turnovers.
    %
    }
    \label{fig:histo}
\end{figure} 
\begin{equation}
p(r)=\sqrt{\frac{2}{\pi}} \frac{r^2 \; \mathrm{exp}(-r^2/a^2)}{a^3},
\label{equation:eq5}
\end{equation}

\noindent 
with a parameter $a=50$~pc. Initially, we distribute  homogeneously 
clumps along the spiral arms; then we add a random noise of $5 \%$ in their polar radius to mimic the fact that \ion{H}{II} regions are not perfectly distributed along the spiral arms and also that the spiral arms themselves are not perfectly delineated in real galaxies. 
We also fix the number of absorbers in a way that the initial (i.e. before adding the random noise) mean separation between them is 5 times their typical size. All  together these conditions result in 981 simulated \ion{H}{II} regions that trace approximately the spiral arms and that they do not overlap each, consistent with the number and distribution of Galactic \ion{H}{II} regions revealed by observational surveys that typically point out hundreds to thousands of these objects (\citealt{anderson2018,wenger2021}, and references therein).

The predicted characteristic frequency $\nu_{\ast}$ can be estimated using its definition, introduced in Sect.~\ref{sec:analysis}, $\nu_{\ast}=\tau_0^{1/2.1} \nu_0$ where the optical depth $\tau_0$ can be computed from the physical properties of the absorbing ionized gas using the following equation \citep{wilson2009}

\begin{equation}
    \tau_0=3.014 \times 10^{-2} \Big( \frac{\rm{T_e}}{\rm K} \Big)^{-3/2} \Big( \frac{\nu_0}{\rm GHz} \Big)^{-2} \, g_{\rm ff} \, \Big( \frac{\rm{EM}}{\rm pc \; cm^{-6}} \Big),
\end{equation}
where $\rm{T_e}$ is the gas electron temperature,
\begin{equation}
    \rm g_{\rm ff} = ln \Big[ 4.955 \times 10^{-2} \Big( \frac{\nu_0}{\rm GHz} \Big) ^{-1} \Big] +1.5\,ln \Big( \frac{T_e}{\rm K} \Big)
\end{equation}
is the Gaunt factor, and 
\begin{equation}
\label{EM}
   {\rm{EM}}=\int_0^L n_{\rm{e}}^2 \;dx 
\end{equation}

\noindent
is the emission measure depending on  the  electron density, $n_{\rm{e}}$,  integrated along the linear length, $L$, of the intervening medium. 
Assuming  a constant electron density, the emission measure is 
$\mathrm{EM}=n_{\mathrm{e}}^2 \, L$. Therefore, we have $\nu_{\ast} \sim n_{\mathrm{e}}^{0.95} L^{0.47} / \rm{T_e}^{0.71}$, i.e. the characteristic frequency depends more strongly on the electron density adopted, less on the electron temperature,  and more weakly on the path $L$. 
We assume constant electron temperature $\rm{T_e}=5000\,~\mathrm{K}$ and  
density 
$n_{\rm e}=2.0~\mathrm{cm}^{-3}$ \citep{Castelletti2021}.

By adopting these values, our model is calibrated to reproduce an average value $\nu_{\ast}\sim$30~MHz observed in the combined sample of \citet{Castelletti2021} plus this work, 
\citet{Kovalenko1994-spec}, and \citet{Kassim1989} (see Fig.~\ref{fig:histo}).  The linear length $L$ is computed as the cumulative sum of all segments along the integral that intersect circular ionized clumps of radius $r$ along the line of sight from the solar position. 
Notice that the NE2001 model cannot predict such characteristic frequencies $\nu_{\ast} \sim 30$~MHz for SNRs that are located nearby (i.e. at heliocentric distances of $\sim 1$~kpc) due to the low electron density assumed ($n_{\mathrm{e}}\sim 0.2$~cm$^{-3}$), which is approximately one order of magnitude lower than that inferred from the SNR turnover spectra by \citet{Castelletti2021}.

\begin{figure}
\includegraphics[width=0.45\textwidth]{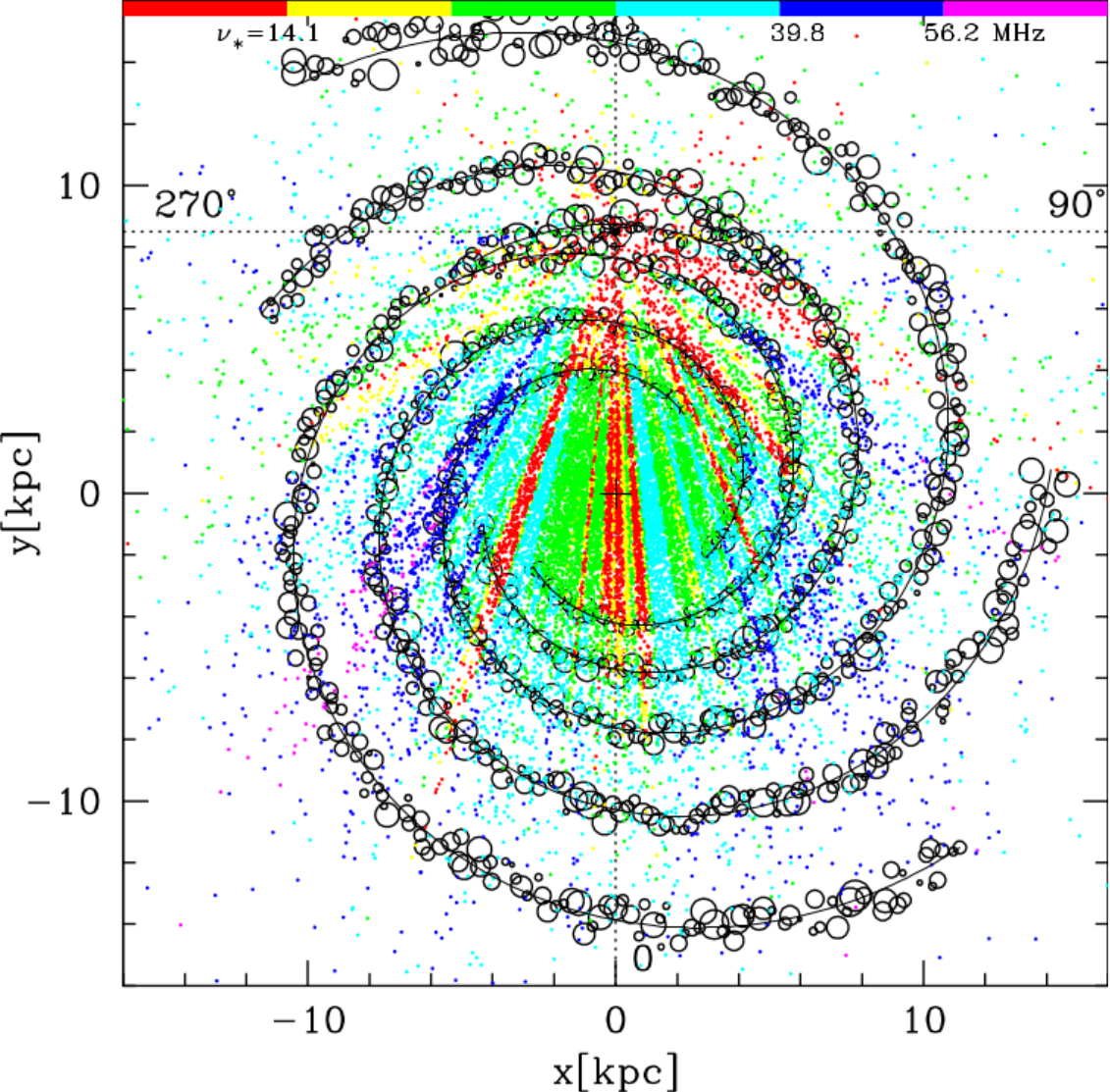}
    \caption{Spatial distribution of an assumed mock sample of  20,000 SNRs (points colored by ${\log}(\nu_{\ast}/\mathrm{MHz})$ using the color bar code). 
    Absorbers (open circles) are distributed along the Galactic  spiral arms taken from the \citet{Hou&Han2014} model (solid black curves). Each absorber is assumed to have a constant low density $n_{\mathrm{e}}=2.0$~cm$^{-3}$, a temperature 
    %
    $\rm{T_e}=5000$~K, and  radius $r$ drawn from a Maxwell-Boltzmann probability density distribution with a median of 50~pc. The scale of each open circle has been expanded by an arbitrary factor \textcolor{violet}{3} in order to make the plot more clear. The central plus symbol indicates the position of the Galactic center, the solar symbol shows the Sun's position, and the dotted lines the Galactic quadrants. The colored points do not show any strong dependence on heliocentric distance or Galactic longitude (see Fig.~\ref{fig:nusdisall} and\ref{fig:nuslonall}).}
    %
    \label{fig:sa}
\end{figure}

The prediction of our model is shown in Fig.~\ref{fig:sa}, where the spatial distribution of the emitters (SNRs) are plotted as colored dots (using the color coding shown in the upper bar), while the absorbers are displayed as black open circles. 
To make the plot more clear the scale of each open circle has been expanded by an arbitrary factor of \textcolor{violet}{3}. 
The observer is assumed to be at Cartesian coordinates (0,8.5)~kpc from the Galactic  centre mimicking the Sun's position. 
Colored points correspond to SNRs that are absorbed by ionized gas along the line of sights. 
In a model of sparse and discrete absorbers, like this one, SNRs with and without spectral turnover are predicted to be mixed, i.e. not spatially segregated as expected in a model with absorbers continuously distributed. 
In a continuous model, SNRs with non-thermal turnovers are expected to be located closer to the observer while those showing turnovers should exhibit characteristic frequencies $\nu_{\ast}$ increasing with distance. This is because the radiation passes through more gas as it travels towards the observer, leading to increased intersection and thus a higher predicted frequency. 
Notice that the asymmetric coil of spiral arms, i.e. that its Galacto-centric distance increases with its polar angle, makes absorption asymmetric between the 1st and 4th and between the 2nd and 3rd quadrants. The strongest absorption is predicted in the fourth quadrant tangential direction (see magenta dots) of the most internal spiral arm.

We conducted a series of tests on our model using different values of the scale-length parameter, $R_d$, for the exponential distribution of SNRs. We also experimented with using a power-law profile, $\Sigma(R)=\Sigma_0(R/R_d)^{\gamma}$. Despite these changes, we found that our results remained consistent and robust.

\begin{figure}
    \includegraphics[width=0.45\textwidth]{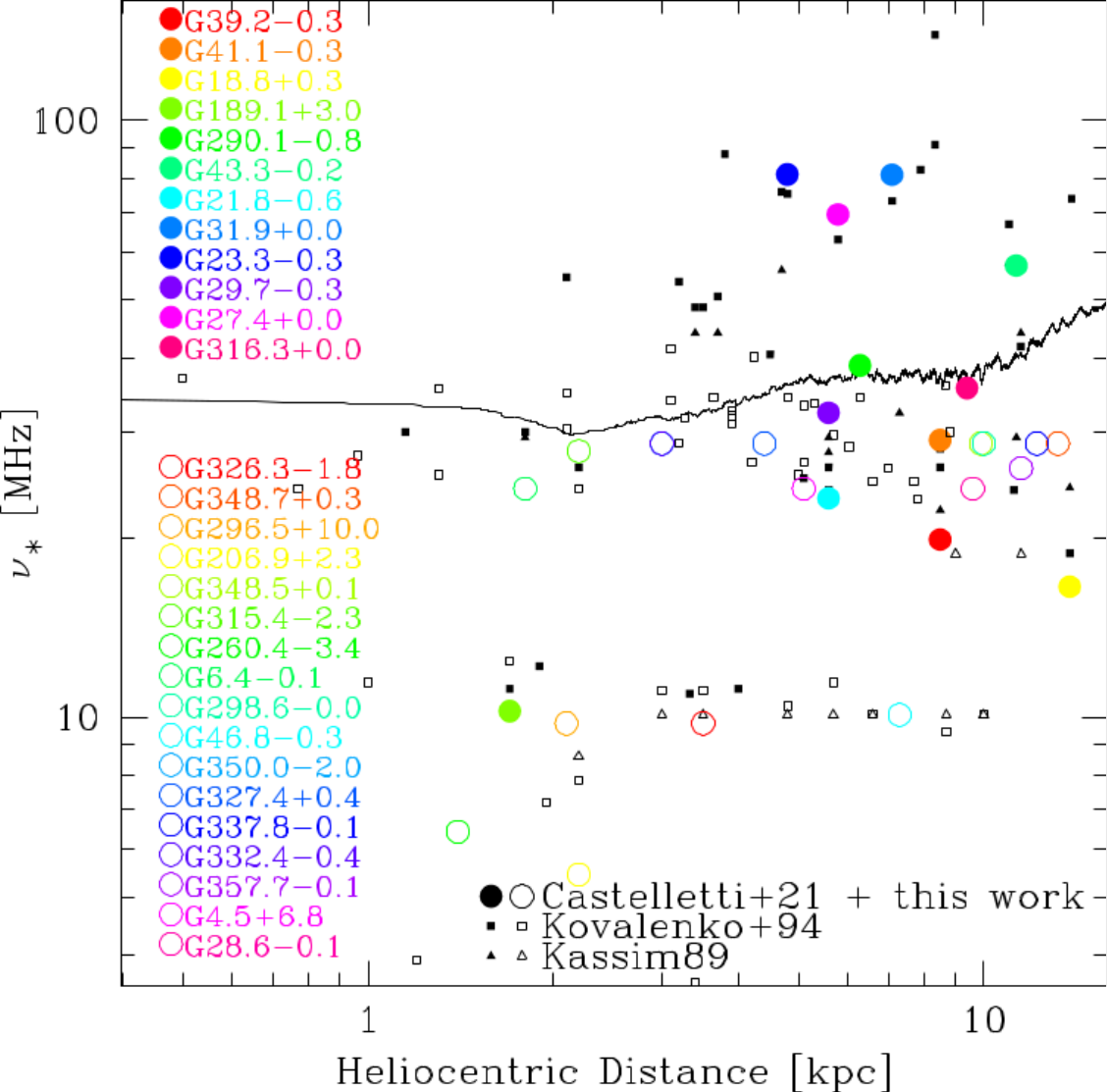}
    \caption{Characteristic frequency $\nu_{\ast}$ as a function of  the heliocentric distance for SNRs with low frequency turnovers (filled symbols)  and  SNRs spectra without turnovers (open symbols). Circles correspond to \citet{Castelletti2021} plus this work data, while squares to  \citet{Kovalenko1994-spec}, and triangles to \citet{Kassim1989}. 
    The solid line corresponds to the median prediction of the 
    model presented in Fig.~\ref{fig:sa} not showing a systematic gradient with heliocentric distance.}
    \label{fig:nusdisall}
\end{figure}

To compare observational results against the predictions of our simple model, in Figs.~\ref{fig:nusdisall} and \ref{fig:nuslonall} we plot the characteristic frequency $\nu_{\ast}$ as a function of the  heliocentric distance and the Galactic longitude, respectively. 
Observational values for spectra with turnovers, indicated by colored filled circles (\citealt{Castelletti2021} plus this work), small black filled squares \citep{Kovalenko1994-spec} and small black filled triangles \citep{Kassim1989}, 
show neither a positive gradient with distance nor a  negative gradient with Galactic  longitude, as expected in a model of continuously distributed absorbing gas. 
Theoretical models featuring discrete absorbers, as depicted in Fig.~\ref{fig:sa} reveal that the values of $\nu_{\ast}$ remain approximately constant. This indicates that absorption is independent of heliocentric distance or Galactic  longitude, as illustrated by the black continuous solid line. 
Besides, we have also included observed SNRs with no measured spectral turnovers (i.e. sources with pure power-law  continuum spectra) with $\nu_{\ast}$ computed as the upper limit allowed by the lowest observed frequency $\nu_{\mathrm{low}}$ (see Sec.~\ref{sec:section2.2}). 
These remnants are located not only quite close, $d \sim 2$~kpc, to the solar position, but also at distances as large as  $d \sim 10$~kpc. Similarly, they can be found either quite close the Galactic center or in the anticenter direction. 

Based on our assumed distribution of \ion{H}{II} regions following the spiral arms, we find that the absorption of our mock SNR population is consistent with the patchy distribution of absorption indicated by observations to date. A much larger sample of low frequency spectra are needed to verify if our model of absorbing gas due to \ion{H}{II} regions is the correct approach.

\section{Conclusions}
\label{sec:conclusions}
We have analyzed a sample of 129 radio continuum spectra of Galactic SNRs from this work and the literature (including \citealt{Castelletti2021}, \citealt{Kovalenko1994-spec}, and \citealt{Kassim1989}). 57 show low frequency turnovers characteristic of extrinsic thermal absorption, while 72 remain straight power laws (special cases of intrinsic absorption by unshocked ejecta within SNRs is not considered in this paper). Since the frequencies sampled vary across the spectra, we take care to interpret the latter as limits on absorption in context of their lowest frequency measurement. 
Our main conclusions are:

\begin{enumerate}
\item We introduced a new parametrization of SNR integrated radio continuum spectra in terms of characteristic frequency $\nu_{\ast}$ and corresponding flux density $S_{\ast}$ independent of arbitrary survey  reference frequencies. $\nu_{\ast}$ 
is directly related both to the power-law emission slope and the turnover frequency  
including the density, temperature, and path through the absorbing medium. 

\item Normalizing the SNR spectra to the frequency $\nu_{\ast}$ and to the flux density $S_{\ast}$, we confirmed with our larger sample that while all SNRs emit energy with independent synchrotron emission  power-law indices, turnovers are all well fit by thermal absorption. 
%


\item For SNRs not showing turnovers we derive an upper limit for $\nu_{\ast}$ by considering the lowest sampled frequency yielding a range of $4~\mathrm{MHz} \lesssim \nu_{\ast} \lesssim 40~\mathrm{MHz}$. This result should be tested and improved by forthcoming surveys pushing SNR spectra to lower frequencies, especially $<$~50~MHz where sensitivity to absorption is markedly increased.

\item Emission ($\alpha$) and absorption ($\nu_{\ast}$) spectral parameters do not show any significant correlation indicating two unrelated, independent processes. For example, absorption beyond the nominal boundaries of the synchrotron emitting plasma seem oblivious to the details and history of the shock acceleration processes within the remnant.
For our three samples of SNR spectra with turnovers 
we found the characteristic frequency ranges 
$10~\mathrm{MHz} \lesssim \nu_{\ast} \lesssim 100~\mathrm{MHz}$.

\begin{figure}
    \includegraphics[width=0.45\textwidth]{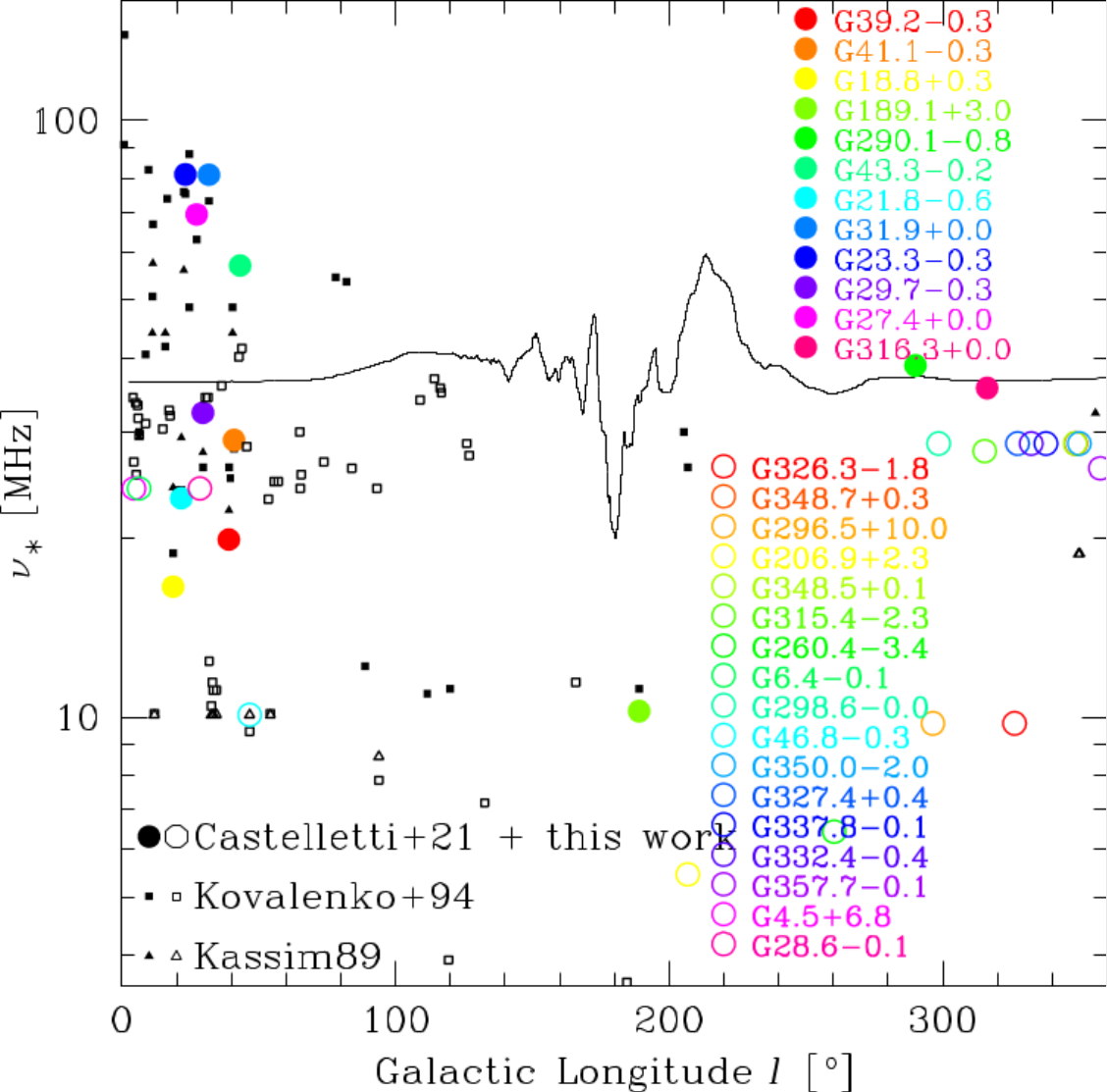}
    \caption{Characteristic frequency $\nu_{\ast}$ as a function of  the Galactic longitude for SNRs with low frequency turnovers (filled symbols)  and SNRs spectra without turnovers (open symbols). Circles correspond to \citet{Castelletti2021} plus this work data, while squares to \citet{Kovalenko1994-spec}, and triangles to \citet{Kassim1989}. 
    The solid line corresponds to the median prediction of the 
    model presented in Fig.~\ref{fig:sa} not showing a strong dependence with the Galactic longitude.}
    \label{fig:nuslonall}
\end{figure}

\item There is no evidence for a correlation between absorption and global Galactic geometric parameters such as heliocentric distance and Galactic longitude, as would be expected if the distribution of the absorbing gas were continuous. 
Our result is also valid for pure power-law SNR spectra. 
This is consistent with \citet{Kassim1989} who inferred that SNRs were being absorbed by an intervening patchy distribution of \ion{H}{II} regions or their envelopes whose distances were unknown. 

\item Modelling based on a distribution of \ion{H}{II} regions tracking Galactic spiral arms produces patchy absorption towards a mock SNR population consistent with the limited observations to date. One prediction of this scheme is that strongly absorbed SNRs should be located preferentially towards the fourth Galactic quadrant relative to the first.

\item Emission measures predicted from NE2001 could not match the observations.  We attribute the discrepancy to the fact that the NE2001 model assumes a nearly continuous distribution of ionized gas, and relies predominantly on a relatively nearby population of PSRs. The NE2001 model predicts a correlation between absorption and distance, particularly for greater distances, that is not observed.
\end{enumerate}

Unfortunately, the incompleteness of current SNR catalogs, and absence of low frequency spectra even for those that are known, preclude making a much more detailed comparison between models and observations. This motivates the completion of deeper, global surveys at both MHz and GHz frequencies, to push all SNR spectra to lower frequencies and to fully sample the  intrinsic properties like density, temperature, size, and distance, of intervening Galactic  thermal gas. Arc-second resolution observations below 100 MHz, as provided by LOFAR and emerging low frequency instruments like LWA and SKA-Low, are critical for discerning the relative radial superposition of SNRs and \ion{H}{II} regions along complex lines of sight. Given readily available kinematic distances to \ion{H}{II} regions, this can offer powerful physical constraints for SNRs whose distances are far more problematic.

\begin{acknowledgements}
This work has been partially supported by the Consejo de
Investigaciones Cient\'ificas y T\'ecnicas de la Rep\'ublica Argentina
(CONICET), the Secretar\'ia de Ciencia y T\'ecnica de la
Universidad Nacional de C\'ordoba (SeCyT), Agencia Nacional de 
Promoci\'on Cient\'ifica y Tecnol\'ogica (PICT 2017-3320, PICT 2019-1600), Argentina. MA acknowledges the hospitality of the Astronomical Observatory of Trieste where part of the work was done in the framework of the European Union LACEGAL program. Basic research in radio astronomy at the Naval Research Laboratory is funded by 6.1 Base funding.  Part of this research was carried
out at the Jet Propulsion Laboratory, California Institute of
Technology, under a contract with the National Aeronautics and Space Administration.

\end{acknowledgements}

\section*{Data availability}

The data underlying this article will be shared on reasonable request to the corresponding author.

\bibliographystyle{aa}
\bibliography{Bibliography}

\begin{appendix}
\section{Fitted parameters of SNRs spectra sample}
\begin{table*}
  \caption{Each group of rows, separated by a horizontal line, corresponds to \cite{Castelletti2021} and this work data, \cite{Kovalenko1994-spec}, and \cite{Kassim1989} sample, respectively. Column 1 is the SNR identification name while columns 2 to 6 are published parameter for Eq.~\ref{equation:eq1} taken from these references. For SNRs G189.1+3.0, G290.1$-$0.8, and G316.3+0.0, we derived the parameters through spectral fits to the datasets constructed for this study. 
  Column 7 and 8 display our computed values  for $\nu_{\ast}$  and $S_{\ast}$, respectively, using $\nu_{\ast}=\tau_0^{1/2.1}\nu_0$ and $S_{\ast}=S_1(\nu_{\ast}/\nu_1)^{\alpha}$ (see Eq.~\ref{equation:eq2}).} 
  
   \label{table:samples}
\centering
   \begin{tabular}{llrlrrrr}
    \hline\hline
\multicolumn{1}{c}{SNR}  & \multicolumn{1}{c}{$\alpha$ } & 
\multicolumn{1}{c}{$\nu_0$} & 
\multicolumn{1}{c}{$\tau_0$} & 
\multicolumn{1}{c}{$\nu_1$} & 
\multicolumn{1}{c}{$S_1$} & 
\multicolumn{1}{c}{$\nu_{\ast}$} & 
\multicolumn{1}{c}{$S_{\ast}$} 
\\
 \multicolumn{1}{c}{} & 
 \multicolumn{1}{c}{} &
 \multicolumn{1}{c}{[MHz]} &
 \multicolumn{1}{c}{} & 
 \multicolumn{1}{c}{[MHz]} & 
 \multicolumn{1}{c}{[Jy]} &
 \multicolumn{1}{c}{[MHz]} &
 \multicolumn{1}{c}{[Jy]} 
\\     
\hline
G18.8+0.3 (Kes~67) & $-0.373$ & 74 & 0.043 & 74 &  79.75 & 16.54 &  139.46
\\
G21.8$-$0.6 (Kes~69)& $-0.505$ & 74 & 0.088 & 74 & 208.00 & 23.26 & 373.16
\\
G23.3$-$0.3 (W41)  & $-0.628$ & 74 & 1.214 & 74 &  347.20 & 81.16 & 327.62 
\\
G27.4+0.0 (Kes~73)& $-0.690$ & 74 & 0.878 & 74 &  34.72 & 69.55 &  36.24 
\\
G29.7$-$0.3 (Kes~75)& $-0.659$ & 74 & 0.176 & 74 &  50.87 & 32.36 &  87.74 
\\
G31.9+0.0 (3C~391)& $-0.521$ & 74 & 1.210 & 74 &  90.70 & 81.03 &  86.51 
\\
G39.2$-$0.3 (3C~396)& $-0.351$ & 74 & 0.063 & 74 &  41.65 & 19.84 &  66.11 
\\
G41.1$-$0.3 (3C~397)& $-0.356$ & 74 & 0.141 & 74 &  75.30 & 29.11 &  104.96 
\\
G43.3$-$0.2 (W49B) & $-0.461$ & 74 & 0.580 & 74 & 111.80 & 57.09 & 125.96
\\
G189.1+3.0 (IC~443) & $-0.388$ & 74 & 0.016 & 74 & 436.90 & 10.24 & 941.26
\\
G290.1$-$0.8 (MSH~11$-$61A) & $-0.421$ & 74 & 0.258 & 74 & 137.80 & 38.82 & 180.80
\\
G316.3+0.0 & $-0.893$ & 74 & 0.215 & 74 & 166.20 & 35.59 & 319.52
\\
\hline
G0.9+0.1          & $-0.64$  & 100 & 0.820 &     2000 &  6.9 &     90.98 &      49.86  
\\
G1.05$-$0.1         & $-0.66$  & 100 & 2.000 &   1700 &     13.0 &   139.10 &           67.83    
\\
G6.4$-$0.1 (W28)     & $-0.42$  & 100 & 0.080 &    450 &  440.0 &     30.04 &        1371.00    
\\
G8.7$-$0.1 (W30)     & $-0.48$  & 100 & 0.150 &     940 &  80.0 &     40.52 &           361.80    
\\
G9.8+0.6          & $-0.56$  & 100 & 0.670 &      740 &  4.1 &     82.64 &           13.99    
\\
G11.2$-$0.3         & $-0.49$  & 100 & 0.240 &     6000 &  9.2 &     50.68 &           95.43    
\\
G11.4$-$0.1         & $-0.46$  & 100 & 0.430 &     1000 &   6.0 &    66.91 &           20.82    
\\
G15.9+0.2         & $-0.63$  & 100 & 0.160 &     1100 &  4.2 &     41.78 &           32.97   
\\
G16.7+0.1         & $-0.53$  & 100 & 0.530 &      960 &   2.9 &    73.91 &           11.29    
\\
G18.8+0.3 (Kes~67)  & $-0.42$  & 100 & 0.030 &     580 &  37.0 &     18.83 &            156.10   
\\
G20.0$-$0.2           & $-0.00$  &  100 & 0.05 &     180 &  10.0 &    24.01 &           10.00
\\
G21.8$-$0.6 (Kes~69)  & $-0.50$  & 100 & 0.050 &     600 &   77.0 &    24.01 &            384.90   
\\
G22.7$-$0.2         & $-0.40$  & 100 & 0.560 &   1850 &    59.0 &    75.87 &            211.70   
\\
G23.3$-$0.3 (W41)    & $-0.48$  & 100 & 0.550 &    190 &   190.0 &    75.23 &           296.40   
\\
G24.7$-$0.6         & $-0.59$  & 100 & 0.760 &      800 &  8.0 &     87.75 &          29.47    
\\
G24.7+0.6         & $-0.14$  & 100 & 0.220 &    1600 &   18.0 &    48.63 &           29.36   
\\
G27.4+0.0         & $-0.71$  & 100 & 0.380 &      560 &   9.4 &   63.08 &                44.30
\\
G29.7$-$0.3 (Kes~75)  & $-0.59$  & 100 & 0.060 &     3100 &  4.4 &     26.19 &            73.56  
\\
G31.9+0.0 (3C~391)  & $-0.51$  & 100 & 0.520 &    2200 &   15.5 &    73.24 &             87.89 
\\
G39.2$-$0.3 (3C~396)  & $-0.48$  & 100 & 0.060 &     840 &   20.0 &    26.19 &            105.70  
\\
G39.7$-$2.0 (W50)    & $-0.50$  & 100 & 0.055 &     760 &   90.0 &    25.13 &              495.00   
\\
G40.5$-$0.5         & $-0.51$  & 100 & 0.220 &     410 &  19.5 &     48.63 &            57.84   
\\
G41.1$-$0.3 (3C~397)  & $-0.46$  & 100 & 0.070 &    10000 &   7.0 &    28.19 &           104.30    
\\
G78.2+2.1 ($\gamma$-Cygni) & $-0.73$  & 100 & 0.280 &    1480 &   40.0 &    54.54 & 2671.00         
\\
G82.2+5.3 (W63) & $-0.63$  & 100 & 0.270 &    1900 &    73.0 &   53.61 &            691.10    
\\
G89.0+4.7 (HB21) & $-0.39$  & 100 & 0.012  &   800 &   240.0   &  12.17 &             1228.00   
\\
G111.7$-$2.1       & $-0.75$  & 100 & 0.010  &  180 &   11500.0   & 10.94  & 93922.52
\\
G120.1+1.4        & $-0.62$  & 100 & 0.010  &  520 &    86.0  &  11.16  & 930.89
\\
G189.1.1$-$3.0      & $-0.42$  & 100 & 0.010  & 700  &   190.0  &  11.16  & 1080.67
\\
G205.5+0.5 (Monoceros) & $-0.47$  & 100 & 0.080 &  1300 &  135.0  &    30.04 &            793.20   
\\
G206.9+2.3 (PKS~0646+06) & $-0.51$  & 100 & 0.060 &     360 &    10.0 &   26.19 &            38.06  
\\
\hline
G6.4$-$0.1 (W28) & $-0.4$   & 30.9 &         0.9 & 408 &         424.0 &       29.39 &    1214.00
\\
G11.2$-$0.3  & $-0.5$   & 30.9 &         2.1 & 408 &          30.0 &       43.99 &      91.36
\\
G11.4$-$0.1  & $-0.4$   & 30.9 &         3.7 & 408 &           7.1 &       57.61 &      15.53
\\
G15.9+0.2  & $-0.7$   & 30.9 &         2.1 & 408 &           8.5 &       43.99 &      40.41
\\
G18.8+0.3 (Kes~67) & $-0.5$   & 30.9 &         0.6 & 408 &          49.0 &       24.23 &     201.10
\\
G21.8$-$0.6 (Kes~69) & $-0.5$   & 30.9 &         0.9 & 408 &          95.0 &       29.39 &     354.00
\\
G22.7$-$0.2  & $-0.2$   & 30.9 &         3.5 & 408 &          65.0 &       56.11 &      96.66
\\
G23.3$-$0.3 (W41) & $-0.4$   & 30.9 &         7.2 & 408 &         119.0 &       79.11 &     229.40
\\
G29.7$-$0.3 (Kes~75) & $-0.6$   & 30.9 &         0.8 & 408 &          15.0 &       27.79 &      75.20
\\
G39.2$-$0.3 (3C~396) &$-0.4$   & 30.9 &         0.5 & 408 &          25.0 &       22.21 &      80.09
\\
G40.5$-$0.5  & $-0.5$   & 30.9 &         2.1 & 408 &          16.0 &       43.99 &      48.73
\\
G41.1$-$0.3 (3C~397) & $-0.4$   & 30.9 &         0.9 & 408 &          39.0 &       29.39 &     111.70
\\
G43.3$-$0.2 (W49) & $-0.4$   & 30.9 &         0.9 & 408 &          46.0 &       29.39 &     131.80
\\
G355.9$-$2.5 & $-0.5$   & 30.9 &         1.1 & 408 &          12.0 &       32.33 &      42.63
\\
    \hline
  \end{tabular}
\end{table*}
\end{appendix}
\end{document}